\documentclass[sigplan,screen]{acmart}

\setcopyright{acmlicensed}
\acmPrice{15.00}
\acmDOI{10.1145/3315507.3330202}
\acmYear{2019}
\copyrightyear{2019}
\acmISBN{978-1-4503-6718-9/19/06}
\acmConference[DBPL '19]{Proceedings of the 17th ACM SIGPLAN International Symposium on Database Programming Languages}{June 23, 2019}{Phoenix, AZ, USA}
\acmBooktitle{Proceedings of the 17th ACM SIGPLAN International Symposium on Database Programming Languages (DBPL '19), June 23, 2019, Phoenix, AZ, USA}

\usepackage[utf8x]{inputenc}
\usepackage{url}
\usepackage{xcolor}
\usepackage{amsmath,amssymb,amsthm,stmaryrd,mathtools}
\usepackage[all]{xy}
\usepackage{hyperref}
\usepackage{bussproofs}
\usepackage{pdftexcmds}
\usepackage{graphicx}

\hypersetup{
    colorlinks=true,
    citecolor=blue,
    linkcolor=blue,
    urlcolor=blue
}

\newcommand{\vect}[1]{\overrightarrow{#1}}

\newcommand{\red}[0]{\mathrel{\triangleright}}

\newcommand{\sem}[1]{\left\llbracket {#1} \right\rrbracket}

\def\orelse{\mathbin{~|~}}
\def\emptymset{\Lbag\Rbag}
\newcommand{\setlit}[1]{\{{#1}\}}
\newcommand{\msetlit}[1]{\Lbag{#1}\Rbag}
\newcommand{\tuple}[1]{\langle{#1}\rangle}
\def\distinct{\delta}
\def\promote{\iota}

\def\kwdo{\mathrm{do}}
\def\setwhere{\mathrm{where}_{\mathsf{set}}}
\def\bagwhere{\mathrm{where}_{\mathsf{bag}}}

\def\kwsel{{\color{blue}\mathtt{SELECT}}}
\def\kwdist{{\color{blue}\mathtt{DISTINCT}}}
\def\kwfrom{{\color{blue}\mathtt{FROM}}}
\def\kwwhere{{\color{blue}\mathtt{WHERE}}}
\def\kwas{{\color{blue}\mathtt{AS}}}
\def\kwunion{{\color{blue}\mathtt{UNION}}}

\def\kwall{{\color{blue}\mathtt{ALL}}}

\def\kwtrue{{\mathsf{true}}}
\def\kwfalse{{\mathsf{false}}}

\def\boolty{\mathbf{B}}

\newcommand{\sfw}[3]{\kwsel~{#1}~\kwfrom~{#2}~\kwwhere~{#3}}
\newcommand{\sdfw}[3]{\kwsel~\kwdist~{#1}~\kwfrom~{#2}~\kwwhere~{#3}}

\DeclareMathOperator{\next}{next}

\newtheorem{theorem}{Theorem}[section]

\title{Mixing set and bag semantics}

\author{Wilmer Ricciotti}
\affiliation{
  \department{LFCS, School of Informatics}              
  \institution{University of Edinburgh}            
}
\email{research@wilmer-ricciotti.net}          
\author{James Cheney}
\affiliation{
  \department{LFCS, School of Informatics}              
  \institution{University of Edinburgh and The Alan Turing Institute}            
}
\email{jcheney@inf.ed.ac.uk}          
\begin{abstract}
  The conservativity theorem for nested relational calculus implies
  that query expressions can freely use nesting and unnesting, yet as
  long as the query result type is a flat relation, these capabilities
  do not lead to an increase in expressiveness over flat relational
  queries.  Moreover, Wong showed how such queries can be translated
  to SQL via a constructive rewriting algorithm.  While this result
  holds for queries over either set or multiset semantics, to the
  best of our knowledge, the questions of conservativity and
  normalization have not been studied for queries that mix set and bag
  collections, or provide duplicate-elimination operations such as
  SQL's $\kwsel~\kwdist$.  In this paper we formalize the problem, and
  present partial progress: specifically, we introduce a calculus with
  both set and multiset collection types, along with natural mappings
  from sets to bags and vice versa, present a set of valid rewrite
  rules for normalizing such queries, and give an inductive characterization of a set of
  queries whose normal forms can be translated to SQL.  We also
  consider examples that do not appear straightforward to translate to
  SQL, illustrating that the relative expressiveness of flat and
  nested queries with mixed set and multiset semantics remains an open
  question.
\end{abstract}

\begin{document}


\maketitle

\section{Introduction}

The nested relational calculus~\cite{BNTW95} provides a principled
foundation for integrating database queries into programming languages.  Wong's conservativity theorem~\cite{wong96jcss} generalized the classic
flat-flat theorem~\cite{ParedaensG92} to show that for any nesting depth $d$, a query
expression over flat input tables returning collections of depth at
most $d$ can be expressed without constructing intermediate results of
nesting depth greater than $d$.  In the special case $d=1$, this
implies the flat-flat theorem, namely that a nested relational query
mapping flat tables to flat tables can be expressed equivalently using
the flat relational calculus.

In addition, Wong's proof technique was constructive, and gave an
easily-implemented terminating rewriting algorithm for normalizing NRC
queries to equivalent flat queries; these normal forms correspond
closely to idiomatic SQL queries and translating from the former to
the latter is straightforward.  The basic approach has been extended
in a number of directions, including to allow for (nonrecursive)
higher-order functions in queries~\cite{Cooper09}, and to allow for
translating queries that return nested results to a bounded number of
flat relational queries~\cite{cheney14sigmod}.  

Normalization-based techniques are used in language-integrated query systems such as
Kleisli~\cite{wong:comprehensions} and Links~\cite{CLWY06}.
Currently, language-integrated query systems such as C\# and
F\#~\cite{meijer:sigmod} support duplicate elimination via a
$\kwdist$ keyword, which is translated to SQL queries in an ad
hoc way, and comes with no guarantees regarding completeness or
expressiveness as far as we know, whereas Database-Supported Haskell
(DSH)~\cite{SIGMOD2015UlrichG} supports duplicate elimination but
gives all operations list semantics and relies on more sophisticated
SQL:1999 features to accomplish this.  Fegaras and Maier~\cite{DBLP:journals/tods/FegarasM00}
propose optimization rules for a nested object-relational calculus with set and bag constructs
but do not consider the problem of conservativity with respect to flat queries.

Wong's proof of conservativity also has the nice property that it
relies on relatively weak properties of collection types.  Thus, it
applies both to set and multiset semantics; if we consider nested
relational queries over sets, then we can translate to SQL queries
using $\kwsel \; \kwdist$ and $\kwunion$ operations that provide set semantics,
while if we consider nested multiset queries we can instead generate
plain $\kwsel$ and $\kwunion \; \kwall$ operations that do not eliminate
duplicates.

SQL itself maintains multiset semantics, but provides several
operations that locally employ set semantics, such as $\kwsel \; \kwdist$
and $\kwunion$. In a database programming context, it seems natural to
consider separate collection types for sets and multisets, so that it
is clear from the type of a query expression whether the multiplicity
matters. The ability to mix set and multiset queries would be beneficial for an accurate implementation of lineage for Links using the technique proposed by Fehrenbach and Cheney~\cite{fehrenbach19}; however, the consequences of this on the
expressiveness of the query language, and the conservativity of nested
set/multiset queries over flat ones, do not appear to be well
understood. This provides concrete motivation for our work.

In this paper we take some first steps towards conservativity and
normal form results for mixed set/multiset queries.  We introduce
$NRC(Set,Bag)$, a straightforward generalization of the nested
relational calculus that contains two collection types (sets and
bags), other standard constructs, and mappings from sets to bags and
vice versa.  The mapping $\promote$ from sets to bags simply coerces a
set to a bag with the same elements, all with multiplicity 1.  The
mapping $\distinct$ from bags to sets performs duplicate elimination:
the set corresponding to a given bag consists of all elements of the
bag with multiplicity $>0$.  We next show that $NRC(Set,Bag)$ can
express conjunctive SQL queries with $\kwsel \; \kwdist$ and
$\kwunion$, illustrating how idiomatic SQL queries can be written in
$NRC(Set,Bag)$.

We then explore the equational rewriting opportunities afforded by
$NRC(Set,Bag)$.  We recapitulate the standard rewriting laws of
collection types in NRC, which apply both to sets and to bags
individually.  We also identify natural properties of $\promote$ and
$\distinct$, particularly relating them to set and bag operations.
The duplicate elimination operation $\distinct$ has several convenient
properties, because (as shown by Lellahi and Tannen~\cite{tannen}) it
is a \emph{monad morphism} from the multiset to set monads. However,
the converse $\promote$ operation has fewer convenient properties.
Nevertheless, $\promote$ and $\distinct$ do form a Galois connection between
the sets and bags over a given type (ordered by the respective
inclusion operations).  Specifically, this means that $\promote$
calculates in some sense the optimal bag among all those that
approximate a given set, and $\distinct$ calculates in some sense the
optimal set among all those that approximate a given bag.  In fact,
this Galois connection is a special case called a Galois insertion,
which means that it satisfies $\distinct\circ \promote = id$, that is,
if we convert a set to a bag and then eliminate duplicates we get back
the original set exactly.

We next discuss the normal forms obtained by applying all possible
rewrite rules until no more subexpressions are reducible.  (We do not
formally explore the termination of this system, but conjecture that
it is terminating.)  We identify normal forms that can be mapped
directly to SQL queries, and give examples for which we do not yet
know a systematic translation.
Nevertheless, we are able to show a weak conservativity result that is
of immediate practical interest: suppose we have queries over flat
inputs and returning flat results.  If we forbid the use of the $\promote$
operation inside bag comprehensions, then the normal form of any
query in this sublanguage is straightforward to translate to SQL.

\section{Language overview}\label{sec:calculus}
We define $NRC(Set, Bag)$ as follows:
\[
\begin{array}{rcl}
  S, T & ::= & A \orelse \tuple{\vect{\ell : T}} \orelse \setlit{T} \orelse \msetlit{T} \\
  L, M, N & ::= & x \orelse c(\vect{M}) \orelse \tuple{\vect{\ell = M}} \orelse M.\ell \\
  & \orelse & \setwhere~M~\kwdo~N \orelse \bagwhere~M~\kwdo~N \\
       & \orelse & \emptyset \orelse \setlit{M} \orelse M \cup N \orelse \bigcup\setlit{M | x \leftarrow N} \\
       & \orelse & \emptymset \orelse \msetlit{M} \orelse M \uplus N \orelse \biguplus \msetlit{M | x \leftarrow N} \\
       & \orelse & \distinct M \orelse \promote M 
\end{array}
\]
Types include atomic types, record types with named fields, sets and bags.
Terms include applied constants, conditional expressions, records with
named fields, and various collection terms (empty, singleton, union,
and comprehension). In this definition, $x$ ranges over variable
names, $c$ over constants, and $\ell$ over record field names. Typing
rules for collections are largely standard. We will allow ourselves to use sequences of generators in comprehensions, which are syntactic sugar for nested comprehensions, e.g.:
\[
\bigcup\setlit{ M | x \leftarrow N, y \leftarrow R } := 
\bigcup\setlit{ \bigcup\setlit{ M | y \leftarrow R } | x \leftarrow N }
\]

We assume an intuitive denotational semantics interpreting these
expressions as finite sets and bags, satisfying the following valid
rules (among others):
\begin{center}
\begin{tabular}{c}
  $\tuple{\ldots, \ell = M, \ldots}.\ell \red M$
  \\
  $\setwhere~\kwtrue~\kwdo~M \red M
  \qquad
  \setwhere~\kwfalse~\kwdo~M \red \emptyset$
  \\
  $\bagwhere~\kwtrue~\kwdo~M \red M
  \qquad
  \bagwhere~\kwfalse~\kwdo~M \red \emptymset$
\end{tabular}

\medskip

\begin{tabular}{r@{$~\red~$}l@{\hspace{.5cm}}r@{$~\red~$}l}
  $\bigcup\setlit{M | x \leftarrow \emptyset}$ & $\emptyset$
  &
  $\biguplus\msetlit{M | x \leftarrow \emptymset}$ & $\emptymset$ 
\end{tabular}
\begin{tabular}{rl}
$\bigcup\setlit{M | x \leftarrow \setlit{N}}$
& $\red M[N/x]$
\\
$\bigcup\setlit{M | x \leftarrow N_1 \cup N_2}$
& $\red \bigcup\setlit{M | x \leftarrow N_1}$
  \\
  & $\quad \cup \bigcup\setlit{M | x \leftarrow N_2}$ 
\end{tabular}
\begin{tabular}{rl}
$\biguplus\msetlit{M | x \leftarrow \msetlit{N}}$
& $\red M[N/x]$
\\
$\biguplus\msetlit{M | x \leftarrow N_1 \uplus N_2}$
& $\red \biguplus\msetlit{M | x \leftarrow N_1}$ 
  \\
  & $\quad \uplus \biguplus\msetlit{M | x \leftarrow N_2}$
\end{tabular}

\medskip

\begin{tabular}{r@{$~\red~$}lr@{$~\red~$}l}
  $\distinct\emptymset$ & $\emptyset$
  & 
    $\promote \emptyset$ & $\emptymset$
  \\
  $\distinct \msetlit{M}$ & $\setlit{M}$
  &
    $\promote \setlit{M}$ & $\msetlit{M}$
  \\
  $\distinct(M \uplus N)$ & $\distinct M \cup \distinct N$
  &
    $\distinct\promote M$ & $M$
\end{tabular}
\end{center}


We can immediately observe the following property about the semantics of 
$\promote$ and $\distinct$:
\begin{proposition}
  For any type $A$, the operations $\promote$ and $\distinct$ form a
  Galois connection between sets and bags of elements of type $A$,
  ordered by subset $\subseteq$ and multiset inclusion $\leq$ orders respectively.
  That is, $\promote(M) \leq N \iff M \subseteq \distinct(N)$. 
  In addition, $\distinct \circ \promote = id$.
\end{proposition}
In addition, we can observe the following relationship between the set
and multiset operations:
\begin{eqnarray*}
M \cup N &=& \distinct(\promote M \uplus \promote N)\\
\bigcup\setlit{M | x \leftarrow N} &=&
\distinct(\biguplus\msetlit{\promote M | x \leftarrow \promote N})
\end{eqnarray*}
Together with identities established earlier, this shows that all of
the set operations in $NRC(Set,Bag)$ can be simulated by $NRC(Bag)$
plus $\distinct$ and $\promote$. This allows us to translate SQL
queries to terms with flat bag type.

\subsection{SQL queries in $NRC(Set,Bag)$}
We can show that $NRC(Set,Bag)$ is sufficiently powerful to express the SQL fragment 
including $\kwsel$ $[\kwdist]$-$\kwfrom$-$\kwwhere$ clauses and $\kwunion \; [\kwall]$. 
Our translation assumes that table names $T$ are interpreted as free variables $x_T$
of a suitable bag type; 
we do not give an explicit translation of SQL terms and conditional expressions, but it is easy 
to express them as combinations of record field projections and $NRC$ constants. 
Also notice that our translation assumes that all terms in the $\kwsel$ clause and 
all subqueries in the $\kwfrom$ clause have been explicitly named using the $\kwas$ keyword.
\begin{gather*}
\sem{T} = x_T
\\
\sem{\sfw{\vect{t~\kwas~\ell}}{\vect{Q~\kwas~y}}{B}}
\\
\qquad = \biguplus\msetlit{\bagwhere~\sem{B}~\msetlit{\tuple{\vect{\ell = \sem{t}}}} | \vect{x \leftarrow \sem{Q}}}
\\
\sem{\sdfw{\vect{t~\kwas~\ell}}{\vect{Q~\kwas~x}}{B}} 
\\
\qquad = \promote\distinct\sem{\sfw{\vect{t~\kwas~\ell}}{\vect{T~\kwas~x}}{B}}
\\
\sem{Q_1~\kwunion~\kwall~Q_2} = \sem{Q_1} \uplus \sem{Q_2}
\\
\sem{Q_1~\kwunion~Q_2} = \promote\distinct\sem{Q_1~\kwunion~\kwall~ Q_2}
\end{gather*}

$\kwsel \; \star$ queries can also be expressed in $NRC(Set,Bag)$ by desugaring them
to named $\kwsel$ queries.

\subsection{Normalization}
The translation of $NRC(Set,Bag)$ into SQL relies on the normalization of queries
into an SQL-like fragment of the formalism by means of a set of
rewrite rules: Fig.~\ref{fig:norm} shows a selection of the rules
(standard rules for set and bag queries are in an appendix). Most of the rules are standard for set and bag queries respectively. Based on the fact that the rewrite rules for set and bag queries, when considered separately, are known to be strongly normalizing and preserve the meaning of expressions, and given that the rules for $\distinct / \promote$ (where the mixing of sets and bags occurs) do not seem to be problematic, we believe our system to be terminating and to preserve the meaning of expressions; we do not know whether it enjoys confluence, but this property is not required
(i.e. we do not require unique normal forms).  

Fortuitously, $\distinct M$ subterms can usually be simplified, and
do not block other rules. On the other hand, $\promote M$ subterms can block
other rewrite rules.  This causes two problems. First, even if the result type of a query is flat,
it might introduce nested structures internally.  For homogeneous set
or bag queries, these nested structures can be normalized away, but in
mixed set--multiset queries,
$\promote M$ can block rewrite rules needed to unnest a nested
set-valued subquery $M$.  We therefore make a simplifying assumption that
$\iota$ and $\delta$ are applied only to flat collections (sets or
multisets of flat records) to avoid this complication.  

Secondly, even with this constraint imposed, the normal form for
bag-queries still allows set-queries $\promote P$ in several positions.
In particular, it is unclear how to unnest set comprehensions within bag comprehensions:
\[
\biguplus\msetlit{M | x \leftarrow \promote\bigcup\setlit{\setlit{N} | y
    \leftarrow P}} \leadsto ???
\]
The normal form for bag-queries must therefore 
allow normalized set-queries $\promote P$ in several positions, particularly 
in comprehension generators $G$. This implies that in a normalized term such as
\[
\biguplus\msetlit{J | x \leftarrow t, y \leftarrow \promote P}
\]
$x$ can actually appear free inside $P$ and be captured by the first generator.
SQL disallows such dependencies between queries in the same $\kwfrom$ clause. 
For example, in the query
%
\[\biguplus\msetlit{J \mid x \leftarrow t, y \leftarrow \promote(P \cup P') }\]
it could be that $z$ appears in $P \cup P'$, but the analogous query
\[
\kwsel~q~\kwfrom~t~\kwas~x, (P~\kwunion~P')~\kwas~y
\]
is not valid SQL if $z$ appears in $P~\kwunion~P'$.

\begin{figure}
\begin{center}\small
\begin{tabular}{rl}
$\bigcup\setlit{M \cup N | x \leftarrow R}$ 
 & $\leadsto$
\\
\multicolumn{2}{r}{$(\bigcup\setlit{M | x \leftarrow R}) \cup (\bigcup\setlit{N | x \leftarrow R})$}
\\
$\bigcup\setlit{M | x \leftarrow N \cup R}$
& $\leadsto \bigcup\setlit{M | x \leftarrow N} \cup \bigcup\setlit{M | x \leftarrow R}$
\\
$\bigcup\setlit{M | y \leftarrow \bigcup\setlit{R | x \leftarrow N }}$
& $\leadsto\bigcup\setlit{ M | x \leftarrow N, y \leftarrow R}$
\\
\\
$\setwhere~M~\kwdo~(N \cup R)$
& $\leadsto$ 
\\
\multicolumn{2}{r}{$(\setwhere~M~\kwdo~N) \cup (\setwhere~M~\kwdo~R)$}
\\
$\setwhere~M~\kwdo \bigcup\setlit{N | x \leftarrow R}$
&  $\leadsto\bigcup\setlit{\setwhere~M~\kwdo~N | x \leftarrow R}$
  \\
&
\\
\multicolumn{2}{c}{$\distinct\emptymset \leadsto  \emptyset
 \qquad \distinct \msetlit{M} \leadsto  \setlit{M}
 \qquad \distinct(M \uplus N) \leadsto  \distinct M \cup \distinct N$}
\\
\multicolumn{2}{c}{$\distinct \biguplus\msetlit{M | x \leftarrow N} \leadsto  
    \bigcup\setlit{\distinct M | x \leftarrow \distinct N}
  \qquad \distinct\promote M \leadsto  M$}
\\
  \multicolumn{2}{c}{$\distinct(\bagwhere~M~\kwdo~N) \leadsto  \setwhere~M~\kwdo~\distinct N$}
\\
\multicolumn{2}{c}{$\promote \emptyset \leadsto \emptymset
  \qquad \promote \setlit{M} \leadsto  \msetlit{M}$}
  \\
  \multicolumn{2}{c}{$\promote(\setwhere~M~\kwdo~N) \leadsto
  \bagwhere~M~\kwdo~\promote N$}
\end{tabular}
\end{center}
\caption{Query normalization (selected rules)}\label{fig:norm}
\end{figure}

The target fragment of NRC for flat queries with type 
$\setlit{\tuple{\vect{\ell:T}}}$ is defined by the following grammar:

\medskip

\begin{tabular}{rl@{\hspace{.5cm}}rl}
$P$ & $::= C_1 \cup \cdots \cup C_n$
&
$F$ & $::= x \orelse \delta x$
\\
$C$ & $::= \bigcup\setlit{H | \vect{z \leftarrow F}}$
&
$R$ & $::= \tuple{\vect{\ell = X}}$
\\
$H$ & $::= \setlit{R} \orelse \setwhere~X~\setlit{R}$
&
$X$ & $::= x.\ell \orelse c(\vect{X})$
\end{tabular}

\medskip

By a similar reasoning, for multiset queries 
we can obtain normal forms described by the following grammar:

\medskip

\begin{tabular}{rl@{\hspace{.5cm}}rl}
$Q$ & $::= D_1 \uplus \cdots \uplus D_n$
&
$J$ & $::= \promote P \orelse \msetlit{R}$
\\
$D$ & $::= \promote P \orelse 
        \biguplus\msetlit{J | \vect{z \leftarrow G}}$
&
& $\orelse ~ \bagwhere~X~\msetlit{R}$
\\
$G$ & $::= x \orelse \promote P$
& 
&
\end{tabular}

\paragraph{Discussion} 
The normal forms $P$ of set queries can be directly translated to equivalent SQL,
replacing $\cup$ with $\kwunion$, comprehensions and $\delta x$ (where
$x$ is a table variable) with
$\kwsel~\kwdist$, and translating NRC record syntax to SQL style.
We can see that unnesting of bag comprehension 
enclosed in a $\distinct$ and used inside a set comprehension can be obtained as 
a derived rule:
\[
\bigcup\setlit{M | x \leftarrow \distinct\biguplus\msetlit{\msetlit{N} | y
    \leftarrow P}} \leadsto \bigcup\setlit{M[N/x] | y \leftarrow \distinct P}
\]

These normal forms suggest a limited form of conservativity which nevertheless appears
practically useful:
\begin{theorem}
  Let $M$ be a query expression whose variables are all of flat
  collection type and whose result is a flat collection type, and
  where $\iota$ and $\delta$ are applied only to flat collections.
  Let $N$ be a normal form of $M$: if there are no occurrences of $\iota$ inside multiset
  comprehensions in $N$, then $N$ can be translated to SQL.
\end{theorem}
Let us note that no rewrite rule can move an $\promote$ into
a multiset comprehension (this would not be the case if we were to add
higher-order functions, however); then, if $M$ has no occurrences of $\promote$ inside bag comprehensions, its normal form also respects this property. We thus know that a sufficient (although not necessary) condition for \emph{un}normalized terms to be translatable to SQL is that they should not contain $\promote$ within a bag comprehension: this can be easily enforced by means of a syntactic check.

\paragraph{Examples} 
An e-commerce company active in several sectors including food and books records
transactions independently for each of its departments, by means of tables
$\mathit{FoodEvents}$ and $\mathit{BookEvents}$ both with attributes
$\mathit{Id}$ and 
\linebreak
$\mathit{EventType}$. The same transaction id can appear
multiple times in the same table to record different events associated with it
(e.g. ``paid'' or ``shipped''), but ids in different tables live in different
namespaces, so that, if an id in $\mathit{FoodEvents}$ and one in
$\mathit{BookEvents}$ are equal, they still refer to different transactions. A
query to collect all the transaction ids of transactions in both departments can
be written in $NRC(Set, Bag)$ as follows:
\[
\begin{array}{l}
\promote\bigcup\setlit{\tuple{\mathit{Id} = f.\mathit{Id}} | f \leftarrow
\mathit{FoodEvents}} 
\\
\qquad \uplus 
\promote\bigcup\setlit{\tuple{\mathit{Id} = b.\mathit{Id}} | b \leftarrow
\mathit{BookEvents}}
\end{array}
\]
or equivalently, in SQL:
\[
\begin{array}{l}
(\kwsel~\kwdist~f.\mathit{Id}~\kwfrom~{\mathit{FoodEvents}~\kwas~f})~\kwunion~\kwall
\\
(\kwsel~\kwdist~b.\mathit{Id}~\kwfrom~{\mathit{BookEvents}~\kwas~b})
\end{array}
\]
The theorem's side conditions are limiting, in that they do exclude certain queries that are straightforward to translate to SQL. For example, the following query performing a join between a bag query and a set query (using table variables $T$, $U$, and $V$) employs $\promote$ inside bag comprehension:
\[
\begin{array}{l}
\biguplus \Lbag \bagwhere~(x.A = y.A)~\msetlit{\tuple{B = x.B, C = y.C}} 
\\
\quad | x \leftarrow T, y \leftarrow \promote(\distinct U \cup \distinct V) \Rbag
\end{array}
\]
We can however easily express the same operation in SQL:
\[
\begin{array}{l}
\kwsel~x.B, y.C
~
\kwfrom~T~\kwas~x, (U~\kwunion~V)~\kwas~y
\\
\kwwhere~x.A=y.A
\end{array}
\]
Notably, this translation works only because $x$ is not used in the
generator for $y$.

It is currently unclear if there exists a general method to normalize
$NRC(Set,Bag)$ queries.  We believe the second constraint
can be lifted by decorrelating set-valued subqueries, but we do not have
insight into how to handle $\promote/\distinct$ applied to nested structures. Let us point out, however, that our result does allow the arbitrary nesting of bag and set queries (including the use of $\promote$) inside a top level set query, because the normal forms of set queries do not contain $\promote$.

\section{Conclusions}

In this short paper we outline initial steps towards conservativity
and normalization results that could provide a solid foundation for
language-integrated query in the presence of mixed set and bag
collections.  The preliminary results in this paper provide criteria that
ensure that mixed set--multiset queries mapping flat inputs to flat
results can be translated to SQL, and which appear to cover many
common cases.  Our results also elucidate the forms of queries
for which this translation is not as straightforward, and resolving
their status will be the focus of future work. 

\paragraph*{Acknowledgments}
This work was supported by ERC Consolidator Grant Skye (grant number
  \grantnum{ERC}{682315}).

{ \bibliographystyle{abbrv}
\bibliography{paper}}

\newpage
\appendix

\section{Type system}
We show here the typing rules for $NRC(Set,Bag)$, which we omitted from Section~\ref{sec:calculus} due to space constraints: the symbol $\boolty$ stands for the Boolean type.

\begin{center}
\AxiomC{$\phantom{A}$}
\UnaryInfC{$\Gamma \vdash \emptyset : \setlit{T}$}
\DisplayProof
\hspace{.5cm}
\AxiomC{$\Gamma \vdash M : T$}
\UnaryInfC{$\Gamma \vdash \setlit{M} : \setlit{T}$}
\DisplayProof

\bigskip

\AxiomC{$\Gamma \vdash M : \setlit{T}$}
\AxiomC{$\Gamma \vdash N : \setlit{T}$}
\BinaryInfC{$\Gamma \vdash M \cup N : \setlit{T}$}
\DisplayProof

\bigskip

\AxiomC{$\Gamma, x:T \vdash M : \setlit{S}$}
\AxiomC{$\Gamma \vdash N : \setlit{T}$}
\BinaryInfC{$\Gamma \vdash \bigcup\setlit{M | x \leftarrow N} : \setlit{S}$}
\DisplayProof

\bigskip

\AxiomC{$\phantom{A}$}
\UnaryInfC{$\Gamma \vdash \emptymset : \msetlit{T}$}
\DisplayProof
\hspace{.5cm}
\AxiomC{$\Gamma \vdash M : T$}
\UnaryInfC{$\Gamma \vdash \msetlit{M} : \msetlit{T}$}
\DisplayProof

\bigskip

\AxiomC{$\Gamma \vdash M : \msetlit{T}$}
\AxiomC{$\Gamma \vdash N : \msetlit{T}$}
\BinaryInfC{$\Gamma \vdash M \uplus N : \msetlit{T}$}
\DisplayProof

\bigskip

\AxiomC{$\Gamma, x:T \vdash M : \msetlit{S}$}
\AxiomC{$\Gamma \vdash N : \msetlit{T}$}
\BinaryInfC{$\Gamma \vdash \biguplus\msetlit{M | x \leftarrow N} : \msetlit{S}$}
\DisplayProof

\bigskip

\AxiomC{$\Gamma \vdash M : \msetlit{T}$}
\UnaryInfC{$\Gamma \vdash \distinct M : \setlit{T}$}
\DisplayProof
\hspace{.5cm}
\AxiomC{$\Gamma \vdash M : \setlit{T}$}
\UnaryInfC{$\Gamma \vdash \promote M : \msetlit{T}$}
\DisplayProof

\bigskip

\AxiomC{$\Gamma \vdash M : \boolty$}
\AxiomC{$\Gamma \vdash N : \setlit{T}$}
\BinaryInfC{$\Gamma \vdash \setwhere~M~\kwdo~N : \setlit{T}$}
\DisplayProof

\bigskip

\AxiomC{$\Gamma \vdash M : \boolty$}
\AxiomC{$\Gamma \vdash N : \msetlit{T}$}
\BinaryInfC{$\Gamma \vdash \bagwhere~M~\kwdo~N : \msetlit{T}$}
\DisplayProof
\end{center}

\begin{figure}
\begin{center}\small
\begin{tabular}{rl}
$\bigcup\setlit{\emptyset | x \leftarrow M}$
& $\leadsto \emptyset$
\\
$\bigcup\setlit{M | x \leftarrow \emptyset}$
& $\leadsto \emptyset$
\\
$\bigcup\setlit{M | x \leftarrow \setlit{N}}$
& $\leadsto M[N/x]$
\\
$\bigcup\setlit{M \cup N | x \leftarrow R}$ 
 & $\leadsto$
\\
\multicolumn{2}{r}{$(\bigcup\setlit{M | x \leftarrow R}) \cup (\bigcup\setlit{N | x \leftarrow R})$}
\\
$\bigcup\setlit{M | x \leftarrow N \cup R}$
& $\leadsto$
\\
\multicolumn{2}{r}{$\bigcup\setlit{M | x \leftarrow N} \cup \bigcup\setlit{M | x \leftarrow R}$}
\\
$\bigcup\setlit{M | y \leftarrow \bigcup\setlit{R | x \leftarrow N }}$
& $\leadsto$
\\
\multicolumn{2}{r}{$\bigcup\setlit{ M | x \leftarrow N, y \leftarrow R}$}
\\
$\bigcup\setlit{M | x \leftarrow \setwhere~N~\kwdo~R}$
& $\leadsto$
\\
\multicolumn{2}{r}{$\bigcup\setlit{\setwhere~N~\kwdo~M | x \leftarrow
  R}$}
\\
&
\\
$\setwhere~\kwtrue~\kwdo~M$
& $\leadsto M$
\\
$\setwhere~\kwfalse~\kwdo~M$
& $\leadsto \emptyset$
\\
$\setwhere~M~\kwdo~(N \cup R)$
& $\leadsto$ 
\\
\multicolumn{2}{r}{$(\setwhere~M~\kwdo~N) \cup (\setwhere~M~\kwdo~R)$}
\\
$\setwhere~M~\kwdo~\emptyset$ 
& $\leadsto \emptyset$
\\
$\setwhere~M~\kwdo~\setwhere~N~\kwdo~R$
& $\leadsto$
\\
\multicolumn{2}{r}{$\setwhere~(M \land N)~\kwdo~R$}
\\
$\setwhere~M~\kwdo \bigcup\setlit{N | x \leftarrow R}$
&  $\leadsto$
\\
\multicolumn{2}{r}{$\bigcup\setlit{\setwhere~M~\kwdo~N | x \leftarrow R}$}
  \\
&
\\
\multicolumn{2}{c}{$\distinct\emptymset \leadsto  \emptyset
 \qquad \distinct \msetlit{M} \leadsto  \setlit{M}
 \qquad \distinct(M \uplus N) \leadsto  \distinct M \cup \distinct N$}
\\
\multicolumn{2}{c}{$\distinct \biguplus\msetlit{M | x \leftarrow N} \leadsto  
    \bigcup\setlit{\distinct M | x \leftarrow \distinct N}
  \qquad \distinct\promote M \leadsto  M$}
\\
  \multicolumn{2}{c}{$\distinct(\bagwhere~M~\kwdo~N) \leadsto  \setwhere~M~\kwdo~\distinct N$}
\end{tabular}
\begin{tabular}{rl}
\\
$\biguplus\msetlit{\emptymset | x \leftarrow M}$
& $\leadsto \emptymset$
\\
$\biguplus\msetlit{M | x \leftarrow \emptymset}$
& $\leadsto \emptymset$
\\
$\biguplus\msetlit{M | x \leftarrow \msetlit{N}}$
& $\leadsto M[N/x]$
\\
$\biguplus\msetlit{M \uplus N | x \leftarrow R}$
 & $\leadsto$
\\
\multicolumn{2}{r}{$(\biguplus\msetlit{M | x \leftarrow R}) \uplus (\biguplus\msetlit{N | x \leftarrow R})$}
\\
$\biguplus\msetlit{M | x \leftarrow N \uplus R}$
& $\leadsto$
\\
\multicolumn{2}{r}{$\biguplus\msetlit{M | x \leftarrow N} \uplus \biguplus\msetlit{M | x \leftarrow R}$}
\\
$\biguplus\msetlit{M | y \leftarrow \biguplus\msetlit{R | x \leftarrow N }}$
& $\leadsto$
\\
\multicolumn{2}{r}{$\biguplus\msetlit{ M | x \leftarrow N, y \leftarrow R}$}
\\
$\biguplus\msetlit{M | x \leftarrow \bagwhere~N~\kwdo~R}$
& $\leadsto$
\\
\multicolumn{2}{r}{$\biguplus\msetlit{\bagwhere~N~\kwdo~M | x \leftarrow R}$}
\\
&
\\
$\bagwhere~\kwtrue~\kwdo~M$
& $\leadsto M$
\\
$\bagwhere~\kwfalse~\kwdo~M$
& $\leadsto \emptymset$
\\
$\bagwhere~M~\kwdo~(N \uplus R)$
& $\leadsto$
\\
\multicolumn{2}{r}{$(\bagwhere~M~\kwdo~N) \cup (\bagwhere~M~\kwdo~R)$}
\\
$\bagwhere~M~\kwdo~\emptymset$ 
& $\leadsto \emptymset$
\\
$\bagwhere~M~\kwdo~\bagwhere~N~\kwdo~R$
& $\leadsto$
\\
\multicolumn{2}{r}{$\bagwhere~(M \land N)~\kwdo~R$}
\\
$\bagwhere~M~\kwdo \biguplus\msetlit{N | x \leftarrow R}$
& $\leadsto$
\\
\multicolumn{2}{r}{$\biguplus\msetlit{\bagwhere~M~\kwdo~N | x \leftarrow R}$}
\\
&
\\
\multicolumn{2}{c}{$\promote \emptyset \leadsto \emptymset
  \qquad \promote \setlit{M} \leadsto  \msetlit{M}$}
  \\
  \multicolumn{2}{c}{$\promote(\setwhere~M~\kwdo~N) \leadsto
  \bagwhere~M~\kwdo~\promote N$}
\\
&
\\
$\tuple{\ldots, \ell = M, \ldots}.\ell \leadsto M$
\end{tabular}
\end{center}
\caption{Query normalization (full)}\label{fig:norm_full}
\end{figure}

\end{document}